\documentclass{ws-procs9x6}

\def\etal {{\it et al.}}

\newcommand{\unit}[1]{~\mathrm{#1}}

\newcommand{\zu}[1]{Fig.~{\ref{#1}}}
\def\rtHz {{\rm \sqrt{Hz}}}

\begin{document}

\title{TESTING LORENTZ INVARIANCE WITH A\\ DOUBLE-PASS OPTICAL RING CAVITY}

\author{Y.\ MICHIMURA,$^{a*}$ N.\ MATSUMOTO,$^a$ N.\ OHMAE,$^b$\\ 
W.\ KOKUYAMA,$^c$ Y.\ ASO,$^a$ M.\ ANDO,$^a$ and K.\ TSUBONO$^a$}

\address{$^a$Department of Physics, University of Tokyo, Bunkyo, Tokyo 113-0033, Japan\\
$^b$Department of Applied Physics, University of Tokyo, Bunkyo, Tokyo 113-8656, Japan\\
$^c$National Metrology Institute of Japan, National Institute of Advanced Industrial Science and Technology (AIST), Tsukuba, Ibaraki 305-8563, Japan\\$^*$E-mail: michimura@granite.phys.s.u-tokyo.ac.jp}

\begin{abstract}
We have developed an apparatus to test Lorentz invariance in the photon sector by measuring the resonant frequency difference between two counterpropagating directions of an asymmetric optical ring cavity using a double-pass configuration. No significant evidence for the violation was found at the level of $\delta c/c \lesssim 10^{-14}$. Details of our apparatus and recent results are presented.
\end{abstract}

\bodymatter

\section{Introduction}
Resonant cavities have been frequently used to test Lorentz invariance in photons. These cavity experiments can be classified into two types, depending on the parity of the cavity used. Even-parity experiments have put limits on even-parity light speed anisotropy to the $\delta c/c \lesssim 10^{-17}$ level\cite{Eisele,Herrmann}.

Our experiment is one of the few odd-parity experiments. We look for a nonzero resonant frequency difference between two counterpropagating directions of an asymmetric optical ring cavity by making use of a double-pass configuration. We have put limits on odd-parity anisotropy to the $\delta c/c \lesssim 10^{-14}$ level\cite{YM}, which is an order of magnitude improvement over previous results from cavity experiments\cite{Herrmann,Baynes}, and is comparable with a result from a Compton scattering experiment\cite{Bocquet}.

\begin{figure}[!t]
\begin{center}
 \psfig{file=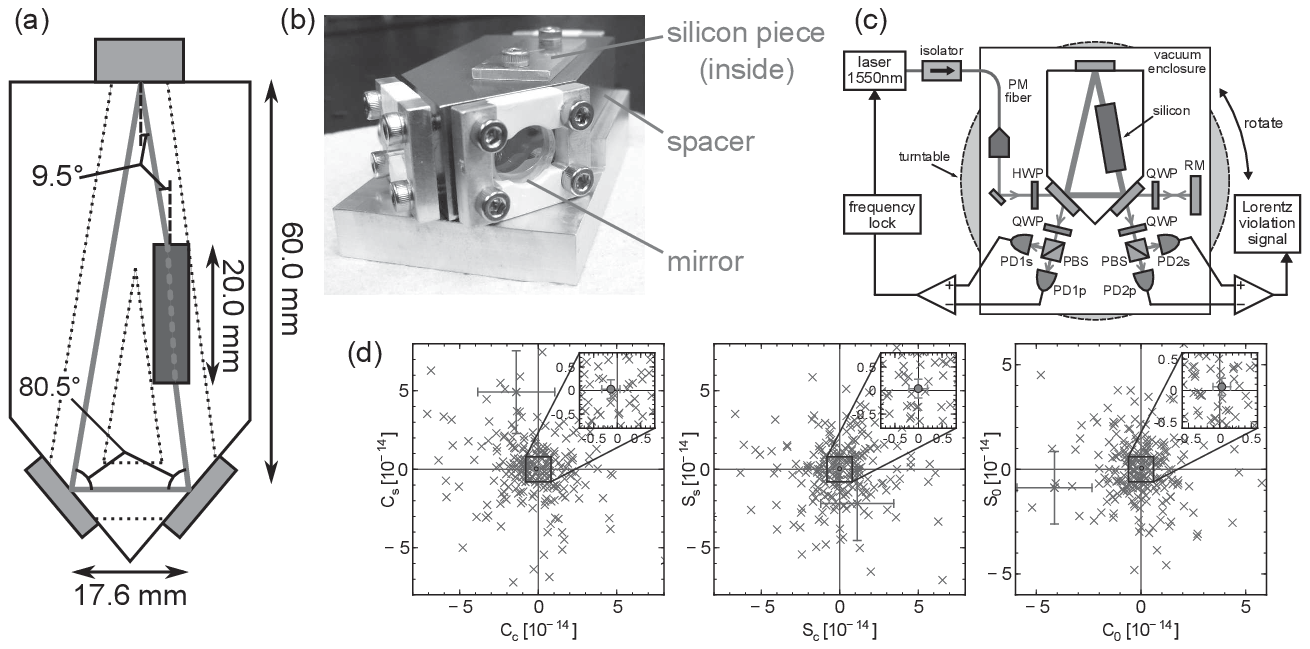,width=4.45in}
\end{center}
\caption{(a) Dimensions and (b) picture of the optical ring cavity. (c) Schematic of the experimental setup. (d) Sidereal modulation amplitudes determined from 1 day data. The mean values and standard errors (shown as a dot in each inset) of all 232 points are $C_{\rm c} = -1.3 \pm 2.0,\ C_{\rm s} = 0.3 \pm 2.0,\ S_{\rm c} = -0.1 \pm 1.9,\ S_{\rm s} = 0.4 \pm 2.0,\ C_{0} = 0.3 \pm 1.4$, and $S_{0} = -0.6 \pm 1.4$ (all values $\times 10^{-15}$).} \label{michi:fig1}
\end{figure}

\section{Experiment}
A simple way to express the odd-parity Lorentz violation is to express the speed of light with $c=1+\alpha \cos{\theta}$, where $\theta$ is the direction of the light propagation with respect to the preferred frame and $\alpha$ is an anisotropy parameter that is zero when there is no violation. From this expression, the resonant frequency shift ($\delta \nu$) of a cavity due to the anisotropy is calculated as
\begin{equation}
  \frac{\delta \nu}{\nu} = \alpha \oint n(\theta) \cos{\theta} r {\rm d} \theta,
\end{equation}
where $n(\theta)$ is the refractive index through the path, and the integral is taken over the round-trip path of the cavity. $\delta \nu$ will be nonzero if $n(\theta)$ changes asymmetrically. Also, since the signs of $\delta \nu$ will be opposite between the round-trip integral taken clockwise and counterclockwise, measuring the resonant frequency difference between two counterpropagating directions gives us a null measurement of the anisotropy parameter $\alpha$. This differential measurement is highly insensitive to environmental disturbances because the effects of cavity length fluctuations are common to both resonances.

Our ring cavity, shown in \zu{michi:fig1}(a) and (b), is a triangular cavity constructed from three half-inch mirrors. These mirrors are mechanically fixed on a spacer made of Super Invar. The spacer has through holes for the optical path. The radius of the holes are $4 \unit{mm}$ and the beam radius inside the cavity is $260 \unit{um}$ at maximum. The spacer also has a hole for placing a silicon piece along one side of the triangle. This silicon piece is rectangular, and its size is $5 \times 10 \times 20 \unit{mm}$. Also, this silicon piece is antireflection coated ($R < 0.5 \%$/surface), and the incident beam to this piece is slightly angled ($\theta_{\rm in}=9.5^{\circ}$) in order to avoid the cross coupling between the counterpropagating beams. Silicon has high transmittance and a large refractive index (measured value $n = 3.69$) at wavelength $\lambda = 1550 \unit{nm}$. The round-trip length of our cavity is $14\unit{cm}$ and the finesse is about 120 for p-polarized light, with the silicon piece inside the cavity.

We use a double-pass configuration for comparing the resonant frequencies between the counterpropagating directions\cite{doublepass} [see \zu{michi:fig1}(c)]. We use a single-frequency laser source with a wavelength of $1550\unit{nm}$ (Koheras AdjustiK C15). Measured relative frequency noise and relative intensity noise without any external stabilization servo was $2 \times 10^{-10}\unit{/\rtHz}$ and $1 \times 10^{-3}\unit{/\rtHz}$ at $0.1\unit{Hz}$ respectively. No external intensity stabilization was employed. The laser beam is fed into the ring cavity in the counterclockwise direction via a polarization maintaining fiber. A collimator (Thorlabs PAF-X-5-C) was used to align and mode-match the incident beam to the ring cavity. The incident beam power to the ring cavity is about 1 mW. 

The frequency of the laser beam is stabilized to the counterclockwise resonance using a piezoelectric actuator attached on the laser cavity. We used the H\"{a}nsh-Coulillaud method~\cite{Hansh} to obtain the error signal for the laser frequency servo. By taking the differential of two PD outputs to obtain an error signal, we reduce the effect of laser intensity fluctuation.

The transmitted light of the counterclockwise beam is then reflected back into the cavity in the clockwise direction by a reflection mirror (RM). We obtain the second error signal, which is proportional to the resonant frequency difference, and in this signal we search for the Lorentz violation. Measured relative frequency noise of the second error signal was $4 \times 10^{-13}\unit{/\rtHz}$ at $0.1\unit{Hz}$. To obtain the second error signal, we again used the H\"{a}nsch-Couillaud method. We did not use Pound-Drever-Hall method because we did not want to introduce spatial mode distortion of the beam from phase modulation.

This double-pass configuration enables a null measurement of the resonant frequency difference with a fairly simple setup. Another way to measure the resonant frequency difference is to inject the laser beam from both directions, stabilize the frequency of each beam to each resonance, and compare the frequency of counterpropagating beams~\cite{Baynes}. However, this configuration needs additional actuation (e.g., acousto-optic modulator) and servo, which could introduce additional noise. Also, stabilizing the frequencies of two counterpropagating beams of a ring cavity has the possibility of lock-in behavior~\cite{lockin}, which is a common effect in ring laser gyroscope.

All the optics are placed in a $30 \times 30 \times 17 \unit{cm}$ vacuum enclosure ($\sim 1\unit{kPa}$) to realize a stable operation. This enclosure is fixed on a turntable together with the laser source, and rotated with a direct drive servo motor (Nikki Denso NMR-CAUIA2A-151A). Positive and reverse rotations of $420^{\circ}$ are repeated alternately in order to avoid twist of the electrical cables. We used thin cables ($0.1 \unit{mm\ dia.}$) for reducing vibrations introduced through the cables. The rotational speed is $\omega_{\rm{rot}}=30^{\circ}$/sec ($f_{\rm{rot}}=0.083$ Hz), and S-curve acceleration and deceleration was used when flipping the sign of the rotations in order to avoid sudden rotational speed change. For data analysis, we only used an interval of $360^{\circ}$ in the middle of each rotation where the rotational speed is constant. Measured rotational speed fluctuation was less than $1\unit{mrad/sec/\rtHz}$ at $0.1\unit{Hz}$.

\section{Recent results and outlook}
Since August 2012, we are continuing the Lorentz violation search. Figure~\ref{michi:fig1}(d) shows recent results from the data taken for 232 days between August 2012 to April 2013. Six sidereal modulation amplitudes defined in Ref.~\refcite{YM} are plotted. Standard errors of each averaged amplitude are $\sim 2 \times 10^{-15}$, which is factor of $\sim 1.4$ improvement over our previous results~\cite{YM}. No deviation from zero by more than $1\sigma$ was found. This means that there is no significant evidence for anisotropy in the speed of light in a sidereal frame at the level of $\delta c/c \lesssim 10^{-14}$.

We are expecting to achieve a year-long run in August 2013. By using the year-long data, we are planning to put first limits on odd-parity higher-order coefficients in SME~\cite{Mewes}.

\section*{Acknowledgments}
We thank Matthew Mewes for useful discussions. This work was supported by Grant-in-Aid for JSPS Fellows number 25$\cdot$10386.


\begin{thebibliography}{x}

\bibitem{Eisele}
Ch.\ Eisele \etal,
Phys.\ Rev.\ Lett.\ {\bf 103}, 090401 (2009).

\bibitem{Herrmann}
S.\ Herrmann \etal,
Phys.\ Rev.\ D {\bf 80}, 105011 (2009).

\bibitem{YM}
Y.\ Michimura \etal,
Phys.\ Rev.\ Lett.\ {\bf 110}, 200401 (2013).

\bibitem{Baynes}
F.N.\ Baynes \etal,
Phys.\ Rev.\ Lett.\ {\bf 108}, 260801 (2012).

\bibitem{Bocquet}
J.-P.\ Bocquet \etal,
Phys.\ Rev.\ Lett.\ {\bf 104}, 241601 (2010).

\bibitem{doublepass}
B.J.\ Cusack \etal,
Class.\ Quantum\ Grav.\ {\bf 19}, 1819 (2002).

\bibitem{Hansh}
T.W.\ H\"{a}nsch and B.\ Couillaud,
Opt.\ Commun.\ {\bf 35}, 441 (1980).

\bibitem{lockin}
F.\ Zarinetchi and S.\ Ezekiel,
Opt.\ Lett.\ {\bf 11}, 401 (1986).

\bibitem{Mewes}
M.\ Mewes,
Phys.\ Rev.\ D {\bf 85}, 116012 (2012).

\end{thebibliography}
\end{document}